\documentclass[12pt,aps]{revtex4}
\usepackage{graphicx}

\begin{document}

\title{Perturbation hydrogen-atom spectrum in deformed space with minimal length}

\author{M.M. Stetsko\footnote{mykola@ktf.franko.lviv.ua}
and V.M. Tkachuk\footnote{tkachuk@ktf.franko.lviv.ua}}
\address{
Department of Theoretical Physics, Ivan Franko National University of Lviv,
12 Drahomanov St., Lviv, UA-79005, Ukraine}

\begin{abstract}
We study energy spectrum for hydrogen atom
with deformed Heisenberg algebra leading to minimal length. We
develop correct perturbation theory free of divergences. It gives
a possibility to calculate analytically in the $3D$ case the
corrections to $s$-levels of hydrogen atom caused by the minimal
length. Comparing our result with experimental data from precision
hydrogen spectroscopy an upper bound for the minimal length is
obtained.
\end{abstract}

\maketitle

\section{Introduction}
Recently a lot of attention has been attracted to the quantum
mechanical problems linked with generalized (modified) commutation
relations. Such an interest was motivated by the works on quantum
gravity and string theory which suggested the existence of a finite
lower bound to the possible resolution of length (minimal length)
\cite{Gross407,Maggiore65,Witten24}. Kempf et al. showed that
minimal length can be obtained from the deformed Heisenberg
algebra \cite{kempf1,kempf2,kempf3,kempf4,kempf5}. Note that for
the first time the deformed algebra leading
to quantized space-time was introduced by Snyder in the relativistic case \cite{Snyder38}.
In D-dimensional case the deformed algebra proposed by Kempf reads

\begin{eqnarray}\label{a2}
&&[X_i,P_j]=i\hbar(\delta_{ij}(1+\beta P^2)+\beta'P_iP_j), \
\ [P_i,P_j]=0,\\ \nonumber
&&[X_i,X_j]=i\hbar\frac{(2\beta-\beta')+(2\beta+\beta')\beta P^2}
{1+ \beta P^2}(P_iX_j-P_jX_i),
\end{eqnarray}
where $\beta, \beta'$ are parameters of deformation. We suppose that these quantities
are positive $\beta, \beta'\geq 0$.
It can be seen that position operators do not commute so we
have a noncommutative space.
From the uncertainty relation it follows that minimal length is
$\hbar\sqrt{\beta+\beta'}$.
Note that in the special case $\beta'=2\beta$  the position operators in linear
approximation over deformation parameters commute,
i.e. $[X_i,X_j]=0$.

The hydrogen atom is one of the simplest quantum system allowing
highly accurate theoretical prediction and is well studied
experimentally offering the most precisely measured quantities
\cite{Karshenboim1}. Therefore, this simple atom has a crucial role
for our understanding of key points of modern physics. Due to the
singularity of the Coulomb potential at the origin this system is
in particular sensitive to whether there is a fundamental minimal
length. There are only a few papers on the investigation of
hydrogen atom in quantum space with minimal length \cite{brau,benczik,Akhoury37}.

Brau \cite{brau} considered the special case of deformation
$\beta'=2\beta$ and in the linear approximation over the deformation
parameters the energy spectrum of hydrogen atom was calculated.
The general case of deformation
$\beta'\neq 2\beta$ was studied in \cite{benczik}.
Using perturbation theory the authors calculated correction to the energy spectrum
of hydrogen atom. But in order to calculate the corrections to $s$-levels for three
dimensional space the authors
were forced to use a numerical method and cut off procedure
due to the appear once of the term  $\sim {1/r^3}$ in the  Hamiltonian in
linear approximation over $\beta, \beta'$.
We would like to emphasize that without cut off procedure this term
leads to divergence the correction to $s$-levels.

Note also paper  \cite{nieto}
where the comparison between the
"space curvature" effects and minimal length effects for the
hydrogen spectrum was made. In \cite{FITYO} a one
dimensional Coulomb problem was solved exactly.

In the present paper we propose the modified perturbation theory free of divergences
which gives a possibility
to calculate the corrections to all energy levels including $s$-levels.
This paper is organized as follows. In the second section we
obtain corrections to the spectrum of $D$-dimensional Coulomb
problem using ordinal perturbation theory.
In the third section we propose modified perturbation theory
and calculate corrections to the energy
of $s$-levels of hydrogen in three dimensional case. And
finally fourth section contains the discussion.

\section{Perturbation of the energy spectrum}

In this section we consider ordinal perturbation theory similarly as in \cite{benczik}
but using another representation of the deformed algebra. This algebra is more convenient
for the development the modified perturbation theory.

We study the eigenvalue problem for hydrogen atom in $D$-dimensional case
\begin{equation}\label{eqn1}
\left(\frac{\textbf{P}^2}{2m}-\frac{e^2}{R}\right)\Psi=E\Psi
\end{equation}
where operators of position $X_i$ and momentum $P_i$ satisfy
deformed commutation relation (\ref{a2}), $R=\sqrt{\Sigma_{i=1}^DX_i^2}$.

We use the following representation that satisfies the
algebra (\ref{a2}) in the first order in $\beta$, $\beta'$
\begin{eqnarray}\label{rep1}
\left\{
\begin{array}{l}
 X_i=x_i+\frac{2\beta-\beta'}{4}\left(x_ip^2+p^2x_i\right),
\\
P_i=p_i+\frac{\beta'}{2}p_ip^2;
\end{array}
\right.
\end{eqnarray}
where $p^2=\Sigma_{k=1}^Dp_k^2$ and operators $x_i$, $p_i$ obey
canonical commutation relations $[x_i, p_j]=i\hbar\delta_{ij}$.
For the undeformed  Heisenberg algebra the
position representation may be taken: $x_i=x_i$,
$p_i=i\hbar\frac{\partial}{\partial x_i}$.

We write Hamiltonian of equation (\ref{eqn1}) using representation
(\ref{rep1}) and taking into account only the first order terms in
$\beta$, $\beta'$
\begin{equation}\label{H1}
H=\frac{p^2}{2m}+\frac{\beta'p^4}{2m}-\frac{e^2}{\sqrt{r^2+\frac{2\beta-\beta'}{2}
\left(r^2p^2+p^2r^2+\hbar^2D\right)}}
\end{equation}
where $r=\sqrt{\Sigma^D_{i=1}x^2_i}$ and $D$ is the dimension of
space.

Expanding the inverse
distance $R^{-1}$ in the series over parameters of deformation
up to the first order $\beta, \beta'$ we have
\begin{equation} \label{Hr3}
H=\frac{p^2}{2m}+\frac{\beta'p^4}{2m}-e^2\left(\frac{1}{r}-\frac{2\beta-\beta'}{4}\left(
\frac{1}{r}p^2+p^2\frac{1}{r}+\frac{\hbar^2(D-1)}{r^3}\right)\right)
\end{equation}
This Hamiltonian contains correction of first order over $\beta, \beta'$
to undeformed hydrogen atom Hamiltonian.

Now we can calculate the corrections $\Delta E^{(1)}_{nl}$ to the
spectrum using the eigenfunctions of undeformed hydrogen atom
\begin{equation}\label{e1}
\Delta
E^{(1)}_{nl}=\frac{e^2\hbar^2}{a^3n^3}\left(\frac{(D-1)(2\beta-\beta')}
{4\overline{l}(\overline{l}+1)(\overline{l}+\frac{1}{2})}+
\frac{2\beta+\beta'}{\overline{l}+\frac{1}{2}}-
\frac{\beta+\beta'}{\overline{n}}\right)
\end{equation}
where $a$ is the Bohr radius and $\overline{n}=n+\frac{D-3}{2}$,
$\overline{l}=l+\frac{D-3}{2}$, $n$ is the principal quantum number and
$l$ is the orbital quantum number.

Expression (\ref{e1}) is in agreement with the results calculated in
the paper \cite{benczik}. It is worth to mention that in the special
case $D=3$ and $l=0$  expression (\ref{e1}) gives divergent
contribution. It is caused  by the term proportional to $1/r^3$ in Hamiltonian
(\ref{Hr3}).

\section{Modified perturbation theory. Corrections to the energy of $s$-levels in three
dimensional case}

In this section we propose a modified perturbation theory which gives a possibility to
overcome the problem of divergence of the corrections to $s$-levels in three dimensional case.
The idea is to use a shifted expansion of inverse distance $R^{-1}$ which does
not contain divergent terms like $1/{r^3}$. So, we rewrite $R$ as follows
\begin{equation}
R=\sqrt{r^2+b^2+\alpha(r^2p^2+p^2r^2+\hbar^2D-\overline{b}^2)}
\end{equation}
where $\alpha=({2\beta-\beta'})/{2}$ and
$b^2=\alpha\overline{b}^2$.
Next we consider the expansion
over $\alpha(r^2p^2+p^2r^2+\hbar^2D-\overline{b}^2)$ in the vicinity of a new point
$r^2+b^2$.
We will choose the introduced parameter $b$ from the condition that
terms proportional to $1/r^3$ are absent in the series.

In the first order over $\alpha$ we can write
\begin{equation} \label{C}
\sqrt{r^2+b^2+\alpha(r^2p^2+p^2r^2+\hbar^2D-\overline{b}^2)}=\sqrt{r^2+b^2}+\alpha\hat{C}.
\end{equation}

Squaring left and right hand side of (\ref{C}) and taking into account only the term of
first order over $\alpha$ we obtain the following equation for
operator $\hat{C}$
\begin{equation}\label{umova2}
r^2p^2+p^2r^2+\hbar^2D-\overline{b}^2=r
\hat{C}+\hat{C}r
\end{equation}

In order to find $\hat{C}$ we write
the left side of (\ref{umova2}) in the following form
\begin{equation}\label{vyraz}
r^2p^2+p^2r^2+\hbar^2D-\overline{b}^2=\frac{1}{2}\left(
r\left(rp^2+p^2r+\frac{A}{r}\right)+\left(rp^2+p^2r+\frac{A}{r}\right)r\right),
\end{equation}
where  parameter $A=\hbar^2(D-1)-\overline{b}^2$.

It is straightforward to show  using (\ref{vyraz}) that
\begin{equation}
\hat{C}=\frac{1}{2}\left(rp^2+p^2r+\frac{A}{r}\right).
\end{equation}

So we have the following expansion for the distance
\begin{equation}\label{R}
R=\sqrt{r^2+b^2}+\frac{\alpha}{2}\left(rp^2+p^2r+\frac{A}{r}\right).
\end{equation}

It is easy to obtain the inverse distance $R^{-1}$ using (\ref{R})
\begin{eqnarray}\label{r1}
\frac{1}{R}=\frac{1}{\sqrt{r^2+b^2}}-\frac{\alpha}{2\sqrt{r^2+b^2}}
\left(rp^2+p^2r+\frac{A}{r}\right)\frac{1}{\sqrt{r^2+b^2}}=
\\ \nonumber
=\frac{1}{\sqrt{r^2+b^2}}-\frac{\alpha}{2}\left(\frac{1}{r}p^2+p^2\frac{1}{r}+\frac{A}{r^3}\right)
.
\end{eqnarray}

The contributions $1/{\sqrt{r^2+b^2}}$ in the second term of
expansion (\ref{r1}) can be replaced with $1/{r}$ in the linear
approximation over $\alpha$. We demand that our expansion does not
contain the terms proportional to $1/{r^3}$ so we conclude that $A=0$, i.e.
\begin{equation}
b=\hbar\sqrt{\alpha(D-1)}.
\end{equation}
It should be noted that (\ref{r1}) takes place under
condition $b^2>0$, i.e. $2\beta>\beta'$.

We rewrite Hamiltonian (\ref{H1}) applying expansion (\ref{r1}) as follows
\begin{equation}\label{H3}
H=\frac{p^2}{2m}+\frac{\beta'p^4}{2m}-e^2\left(\frac{1}{\sqrt{r^2+b^2}}-\frac{2\beta-\beta'}{4}
\left(\frac{1}{r}p^2+p^2\frac{1}{r}\right)\right)=H_0+V,
\end{equation}
where $H_0$ is the Hamiltonian of undeformed hydrogen atom and perturbation
caused by deformation is
\begin{equation}
V=\frac{\beta'p^4}{2m}-e^2\left(\frac{1}{\sqrt{r^2+b^2}}-\frac{1}{r}\right.
\left.-\frac{2\beta-\beta'}{4}\left(\frac{1}{r}p^2+p^2\frac{1}{r}\right)\right).
\end{equation}

Now we calculate the correction to the ground state energy of hydrogen
atom caused by perturbation $V$. First let us consider correction which is connected with
$1/{\sqrt{r^2+b^2}}-1/r$. We have
\begin{eqnarray}\label{popr1}
\left\langle\Psi_{1s}\left|\frac{1}{\sqrt{r^2+b^2}}\right|\Psi_{1s}\right\rangle
= \frac{4}{a^3}\left(\pi\right.
ab\left(H_1\left(\frac{2b}{a}\right)-Y_1\left(\frac{2b}{a}\right)\right)-
\\
\\ -\frac{\pi b^2}{2}
\left.\left(H_0\left(\frac{2b}{a}\right)-Y_0\left(\frac{2b}{a}\right)\right)\right).
\end{eqnarray}
where $H$ and $Y$ are the Struve and the Bessel functions,
respectively \cite{Abramowitz830}.
Then up to first order over $\alpha$ (or $b^2$)
\begin{equation}\label{popr2}
\left\langle \Psi_{1s}\left|
\frac{1}{\sqrt{r^2+b^2}}-\frac{1}{r}\right|\Psi_{1s}\right\rangle=
\frac{2b^2}{a^3}\left(\ln{\frac{b}{a}}+\gamma+\frac{1}{2}\right).
\end{equation}
where $\gamma=0.57721...$ is the Euler constant.

It is easy to calculate the contributions caused by the
terms $\frac{1}{r}p^2+p^2\frac{1}{r}$ and $p^4$. As a result
the correction to the ground state energy reads
\begin{equation}\label{d3l0}
\Delta E^{(1)}_{1s}=\langle\Psi_{1s}|V|\Psi_{1s}\rangle=
\frac{e^2\hbar^2}{a^3}\left(3\beta+\beta'-(2\beta-\beta')
\left(\ln{\frac{\hbar^2(2\beta-\beta')}{a^2}}+2\gamma+1\right)\right).
\end{equation}

We also calculate the correction to the $2s$-level
\begin{equation}
\Delta
E^{(1)}_{2s}=\langle\Psi_{2s}|V|\Psi_{2s}\rangle=\frac{e^2\hbar^2}{8a^3}\left(
\frac{7\beta+3\beta'}{2}-(2\beta-\beta')\left(\ln{\frac{\hbar^2(2\beta-\beta')}{4a^2}}+
2\gamma+\frac{5}{2}
\right)\right).
\end{equation}
Note that in special case $2\beta=\beta'$ these results for energy levels
reproduce the results of Brau \cite{brau}.

Similarly as in  \cite{benczik} we introduce two
parameters $\xi={\Delta x_{\rm min}}/{a}$ and
$\eta={\beta}/{(\beta+\beta')}$ instead of $\beta$ and $\beta'$,
where the minimal
length $\Delta x_{\rm min}=\hbar\sqrt{\beta+\beta'}$.
We have already noticed that calculated corrections
to the $s$-levels take place under conditions $2\beta-\beta'\geq
0$. The conditions for the parameters $\beta$, $\beta'$ restrict
the domain of variation for the parameter $\eta$. It is easy to
verify that $\frac{1}{3}\leq \eta\leq 1$. We rewrite the
correction for the $1s$-level using the parameters $\eta$ and
$\xi$:
\begin{equation}\label{e0}
\Delta E^{(1)}_{1s}=\frac{e^2}{a}
\xi^2\left(2\eta+1-(3\eta-1)\left(\ln{\xi^2(3\eta-1)}
+2\gamma+1\right)\right).
\end{equation}
The correction to the $2s$-level as the function of
parameters $\xi$ and $\eta$ reads
\begin{equation}
\Delta
E^{(1)}_{2s}=\frac{e^2}{8a}\xi^2\left(\frac{1}{2}(4\eta+3)-(3\eta-1)\right.
\left.\left(\ln
\frac{\xi^2(3\eta-1)}{4}+2\gamma+\frac{5}{2}\right)\right)
\end{equation}

\begin{figure}[h!!]
  \centerline{\includegraphics[scale=0.69,clip]{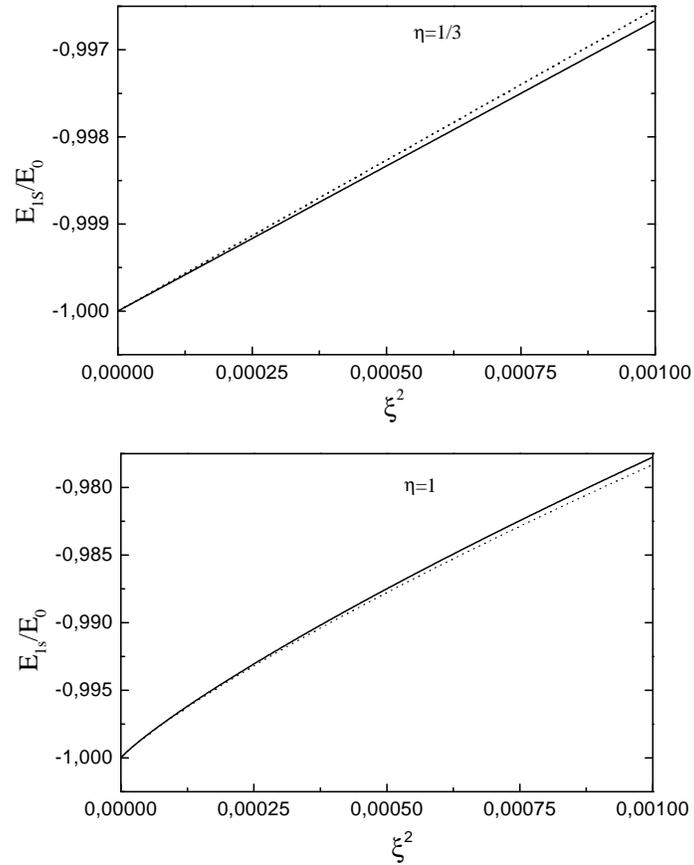}}
  \caption{Comparison of different results for the energy of $1s$ states with $\eta=1/3$
  (upper graph) and $\eta=1$ (lower graph).
  Solid lines represent our results and dotted lines correspond to the
  results of Benczik \cite{benczik}.  }\label{rys2}
\end{figure}

\begin{figure}[h!!]
  \centerline{\includegraphics[scale=0.6,clip]{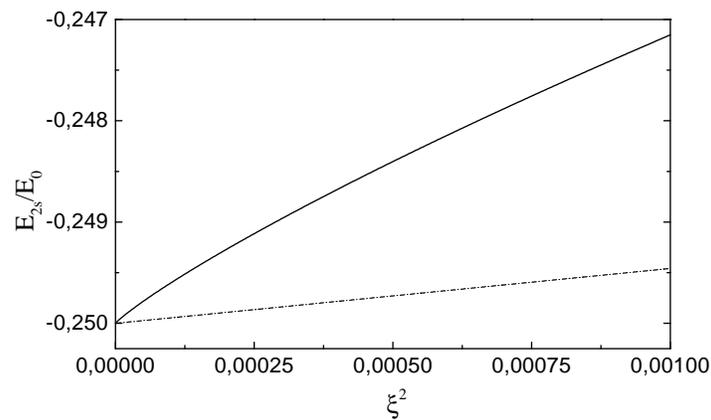}}
  \caption{Energy of $2s$-level.
  Solid line shows the result when $\eta=1$ and
  dash-dotted line corresponds the parameter $\eta={1}/{3}$.  }\label{rysunok}
\end{figure}

Energy of
$1s$-level and $2s$-level as a function of parameter $\xi$ for the fixed value
$\eta$ is presented in Fig.\ref{rys2} and Fig.\ref{rysunok} respectively.
Unit of energy in these figures is the absolute value of the ground state energy of
undeformed hydrogen atom
$E_0=e^2/2a$.

Finally we can consider constraints on the minimal length. As it
was noted \cite{benczik}, the best estimation of the minimal length
can be obtained by including the contributions of the minimal
length effects in the Lamb shift. The experimental Lamb shift for
the $1s$-level of hydrogen atom $L^{\rm
exprt}_{1s}=8172.837(22)$~MHz \cite{schwob} is larger than the best
theoretically obtained $L^{\rm theor}_{1s}=8172.731(40)$~MHz
\cite{mallampalli}.
Assuming similarly as in \cite{benczik} that the discrepancy between experimental
and theoretical values $ L^{\rm
exprt}_{1s}-L^{\rm theor}_{1s}$ is entirely attributed to
the minimal length correction $\Delta E^{(1)}_{1s}$
we find upper bound for the minimal length.

\begin{figure}[h!!]
\centerline{\includegraphics[scale=0.65,clip]{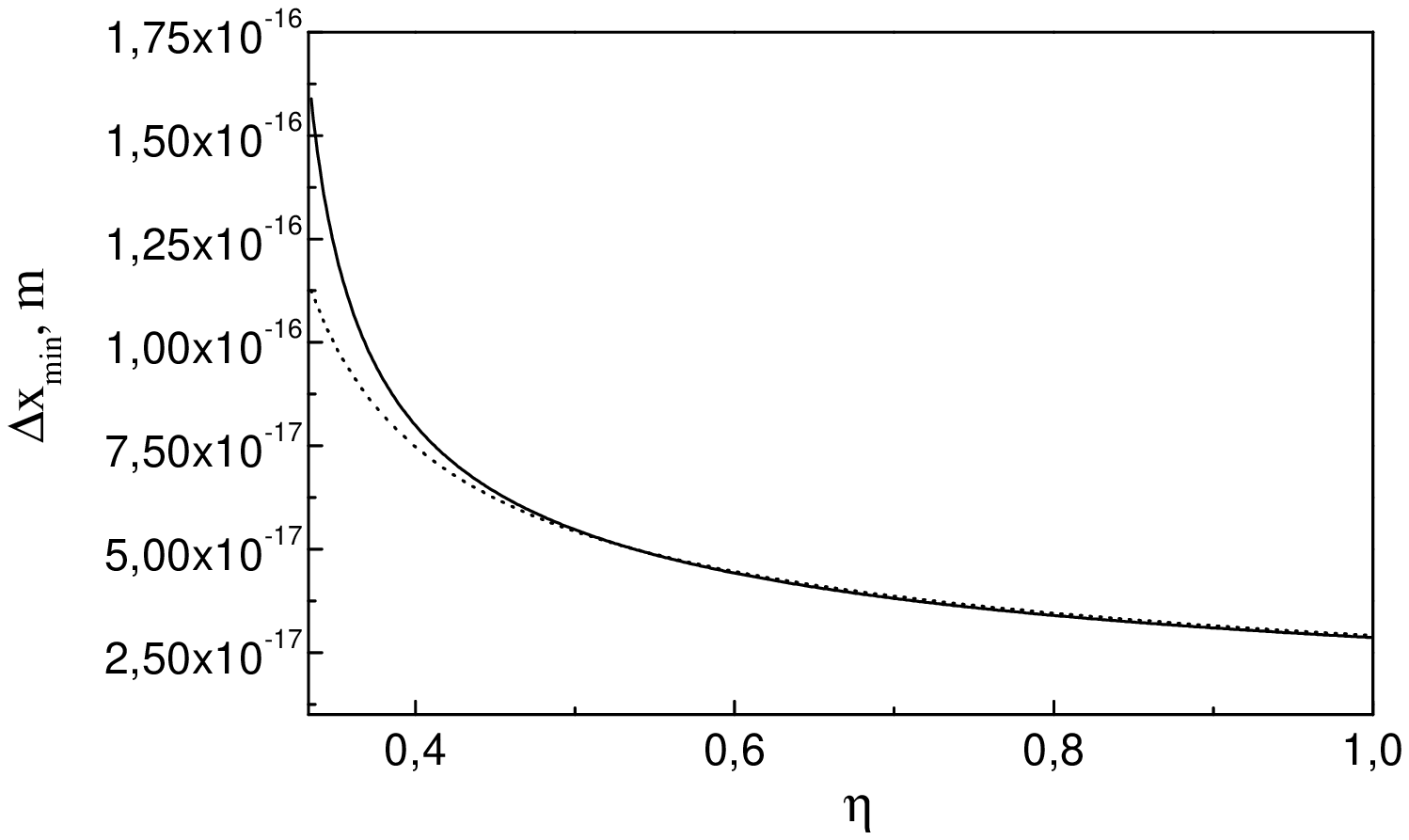}}
\caption{Constraints on the minimal length as a function of
dimensionless parameter $\eta$. Solid line represents the results
obtained here and dotted line shows the results of
\cite{benczik}.}\label{rys3}
\end{figure}

So having relation (\ref{e0}) we can estimate the minimal length
as a function of parameter $\eta$. The limit values of the minimal
length are $\Delta x_{\rm min}=1.64 \cdot 10^{-16}$~m for
$\eta={1}/{3}$ and $\Delta x_{\rm min}=2.86\cdot
10^{-17}$~m for $\eta=1$. Let us compare our results with the
estimation obtained in \cite{benczik} where it was shown that for
$\eta=\frac{1}{3}$ the minimal length $(\Delta
x_{\rm min})^{-1}=1.75$~GeV or $\Delta x_{\rm min}=1.13\cdot 10^{-16}$~m and for
$\eta=1$ the minimal length $(\Delta x_{\rm min})^{-1}=6.87$~GeV or
equivalently $\Delta x_{\rm min}=2.87\cdot 10^{-17}$~m,
where $1{\rm m}=(\hbar c/e)({\rm eV})^{-1}$.
The comparison of
our results for the minimal length with results obtained in
\cite{benczik} for all values of parameter $\eta$ is shown in
Fig.\ref{rys3}.

As on can see our result for energy levels as well as for
estimation of the minimal length are in good agreement with the result
obtained in the paper \cite{benczik},
especially for the $\eta=1$. The discrepancy between
these two estimations are caused by the use different methods of
computation. Namely, using modified perturbation theory we have analytical
expressions for $s$-energy levels. The authors of \cite{benczik}
in order to calculate correction to $s$-levels in three dimensional case used
cut off procedure and numerical calculation. More detailed
comparison is made in Discussion.

\section{Discussion}

We study hydrogen atom in the space with deformed Heisenberg algebra leading to
non-zero minimal length.
The ordinary perturbation theory proposed in \cite{benczik} leads to the term proportional
to $1/r^3$ in perturbation operator.
This term gives a divergent contribution to energy of
$s$-levels in three dimensional case. Therefore, the authors of \cite{benczik} were
forced to use the cut off the expectation value integral $\langle 1/r^3\rangle$
at some point.
In order to find the free parameter appeared as a result
of this procedure the author used the numerical calculation.

We construct a modified perturbation theory free of divergences where instead of $b^2/r^3$
we have $(1/r-1/\sqrt{r^2+b^2})$ with $b=\hbar\sqrt{(D-1)({2\beta-\beta'})/{2}}$.
It gives us a possibility to calculate analytically corrections to energy levels
caused by deformation including $s$-levels in three dimensional case.
Comparing our result with experimental data from precision
hydrogen spectroscopy we find that the upper bound for the minimal length is
of order $10^{-16}$~m, $10^{-17}$~m.
Our results for energy levels as well as the result for
estimation of the minimal length are in a good agreement with the results
obtained in the paper \cite{benczik} with the help of numerical calculation
and cut off procedure.
Note also that for the case $\eta=1/3$ ($2\beta=\beta'$) our results
for energy levels reproduce the result of Brau \cite{brau}, as it must be.

Finally, we would like to draw attention to the problem which is ignored in present
paper as well as in the papers of other authors. Considering
hydrogen atom in deformed space we suppose that Coulomb potential is the same as in non
deformed space. In fact, in the deformed-space the Coulomb potential
as a potential of a point charge might be corrected.
This will lead to additional corrections to the energy spectrum
of hydrogen atom. It is an interesting problem which worth separate investigations.

\end{document}